\documentclass[twocolumn]{aastex631}
\usepackage{amsmath,amssymb}%
\usepackage{rotating}

\newcommand{\ra}[4]{${#1}^{\rm h}{#2}^{\rm m}{#3}\fs{#4}$}
\newcommand{\dec}[4]{${#1}\arcdeg{#2}\arcmin{#3}\farcs{#4}$}

\received{}
\revised{}
\accepted{}

\shorttitle{Millinovae: A New Class of Transient Supersoft X-ray Sources}
\shortauthors{Mr\'oz et al.}

\graphicspath{{./}}

\begin{document}

\title{Millinovae: A New Class of Transient Supersoft X-ray Sources without a Classical Nova Eruption}

\author[0000-0001-7016-1692]{Przemek Mr\'oz}
\affiliation{Astronomical Observatory, University of Warsaw, Al. Ujazdowskie 4, 00-478 Warszawa, Poland}

\author[0009-0006-3777-6381]{Krzysztof Kr\'ol}
\affiliation{Astronomical Observatory, University of Warsaw, Al. Ujazdowskie 4, 00-478 Warszawa, Poland}

\author[0000-0002-9904-3582]{H\'el\`ene Szegedi}
\affiliation{Department of Physics, University of the Free State, 205 Nelson Mandela Drive, Bloemfontein, 9300, South Africa}

\author{Philip Charles}
\affiliation{Department of Physics and Astronomy, University of Southampton, Southampton, Hampshire SO17 1BJ, UK}
\affiliation{Department of Physics, University of the Free State, 205 Nelson Mandela Drive, Bloemfontein, 9300, South Africa}

\author[0000-0001-5624-2613]{Kim L. Page}
\affiliation{School of Physics and Astronomy, University of Leicester, Leicester LE1 7RH, UK}

\author[0000-0001-5207-5619]{Andrzej Udalski}
\affiliation{Astronomical Observatory, University of Warsaw, Al. Ujazdowskie 4, 00-478 Warszawa, Poland}

\author[0000-0002-7004-9956]{David A.H. Buckley}
\affiliation{South African Astronomical Observatory, PO Box 9, Observatory Road, Observatory 7935, South Africa}
\affiliation{Department of Astronomy, University of Cape Town, Private Bag X3, Rondebosch 7701, South Africa}
\affiliation{Department of Physics, University of the Free State, 205 Nelson Mandela Drive, Bloemfontein, 9300, South Africa}

\author{Gulab Dewangan}
\affiliation{Inter-University Centre for Astronomy \& Astrophysics, Post Bag 4, Ganeshkhind, Pune, Maharashtra 411007, India}

\author{Pieter Meintjes}
\affiliation{Department of Physics, University of the Free State, 205 Nelson Mandela Drive, Bloemfontein, 9300, South Africa}

\author[0000-0002-0548-8995]{Micha\l{} K. Szyma\'nski}
\affiliation{Astronomical Observatory, University of Warsaw, Al. Ujazdowskie 4, 00-478 Warszawa, Poland}

\author[0000-0002-7777-0842]{Igor Soszy\'nski}
\affiliation{Astronomical Observatory, University of Warsaw, Al. Ujazdowskie 4, 00-478 Warszawa, Poland}

\author[0000-0002-2339-5899]{Pawe\l{} Pietrukowicz}
\affiliation{Astronomical Observatory, University of Warsaw, Al. Ujazdowskie 4, 00-478 Warszawa, Poland}

\author[0000-0003-4084-880X]{Szymon Koz\l{}owski}
\affiliation{Astronomical Observatory, University of Warsaw, Al. Ujazdowskie 4, 00-478 Warszawa, Poland}

\author[0000-0002-9245-6368]{Rados\l{}aw Poleski}
\affiliation{Astronomical Observatory, University of Warsaw, Al. Ujazdowskie 4, 00-478 Warszawa, Poland}

\author[0000-0002-2335-1730]{Jan Skowron}
\affiliation{Astronomical Observatory, University of Warsaw, Al. Ujazdowskie 4, 00-478 Warszawa, Poland}

\author[0000-0001-6364-408X]{Krzysztof Ulaczyk}
\affiliation{Department of Physics, University of Warwick, Coventry CV4 7 AL, UK}
\affiliation{Astronomical Observatory, University of Warsaw, Al. Ujazdowskie 4, 00-478 Warszawa, Poland}

\author[0000-0002-1650-1518]{Mariusz Gromadzki}
\affiliation{Astronomical Observatory, University of Warsaw, Al. Ujazdowskie 4, 00-478 Warszawa, Poland}

\author[0000-0002-9326-9329]{Krzysztof Rybicki}
\affiliation{Department of Particle Physics and Astrophysics, Weizmann Institute of Science, Rehovot 76100, Israel}
\affiliation{Astronomical Observatory, University of Warsaw, Al. Ujazdowskie 4, 00-478 Warszawa, Poland}

\author[0000-0002-6212-7221]{Patryk Iwanek}
\affiliation{Astronomical Observatory, University of Warsaw, Al. Ujazdowskie 4, 00-478 Warszawa, Poland}

\author[0000-0002-3051-274X]{Marcin Wrona}
\affiliation{Department of Astrophysics and Planetary Sciences, Villanova University, 800 Lancaster Avenue, Villanova, PA 19085, USA}
\affiliation{Astronomical Observatory, University of Warsaw, Al. Ujazdowskie 4, 00-478 Warszawa, Poland}

\author{Mateusz J. Mr\'oz}
\affiliation{Astronomical Observatory, University of Warsaw, Al. Ujazdowskie 4, 00-478 Warszawa, Poland}

\begin{abstract}
Some accreting binary systems containing a white dwarf (such as classical novae or persistent supersoft sources) are seen to emit low-energy X-rays with temperatures of $\sim$10$^6$\,K and luminosities exceeding $10^{35}$\,erg\,s$^{-1}$. These X-rays are thought to originate from nuclear burning on the white dwarf surface, either caused by a thermonuclear runaway (classical novae) or a high mass-accretion rate that sustains steady nuclear burning (persistent sources). The discovery of transient supersoft X-rays from ASASSN-16oh challenged these ideas, as no clear signatures of mass ejection indicative of a classical nova eruption were detected, and the origin of these X-rays remains controversial. It was unclear whether this star was one of a kind or representative of a larger, as yet undiscovered, group. Here, we present the discovery of 29 stars located in the direction of the Magellanic Clouds exhibiting long-duration, symmetrical optical outbursts similar to that seen in ASASSN-16oh. We observed one of these objects during an optical outburst and found it to be emitting transient supersoft X-rays, while no signatures of mass ejection (indicative of a classical nova eruption) were detected. We therefore propose that these objects form a homogeneous group of transient supersoft X-ray sources, which we dub ``millinovae'' because their optical luminosities are approximately a thousand times fainter than those of ordinary classical novae.
\end{abstract}


\section{Introduction}

A classical nova eruption occurs on the surface of a white dwarf that is accreting matter from its nondegenerate companion via an accretion disk. A layer of fresh, usually hydrogen-rich gas accumulates on the white dwarf surface and becomes degenerate. The gas is heated, and when hydrogen ultimately ignites, a thermonuclear runaway ejects about $10^{-5}\!-\!10^{-4}\,M_{\odot}$ of matter from the system with velocities from hundreds to thousands of km\,s$^{-1} $\citep{Bode_Evans_2008}. As the ejecta expands and becomes optically thin, the exposed white dwarf surface is observed to emit supersoft X-rays due to stable hydrogen burning of the leftover gas \citep{Osborne2015,page2022}. The maximum X-ray luminosities of classical novae are typically $10^{36}\!-\!10^{38}$\,erg\,s$^{-1}$ \citep[e.g.,][]{henze2014}. In the optical passbands, they normally reach absolute magnitudes from $-6$ to $-10$ \citep{Bode_Evans_2008}.

\begin{figure*}[htb!]
\centering
\includegraphics[width=.9\textwidth]{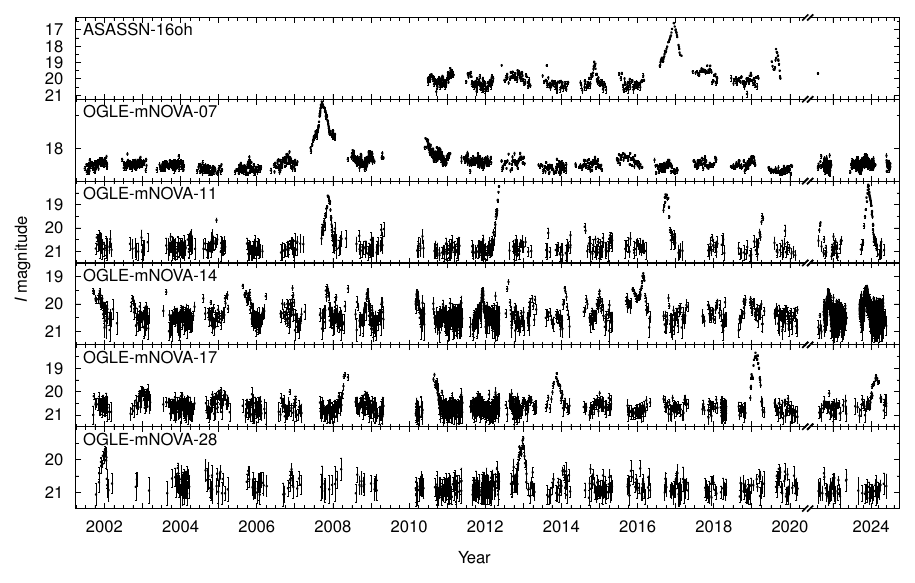}
\caption{Example light curves of millinovae. The upper panel shows the prototype, ASASSN-16oh. The $y$-axis ticks are every 0.5~mag. There is a gap in the data between 2020.4 and 2022.4.}
\label{fig:light_curves}
\end{figure*}

The accretion disk surrounding the white dwarf is subject to a thermal instability that can lead to dwarf nova outbursts. They occur when gas in the disk reaches a critical temperature, and its viscosity abruptly changes, increasing the mass accretion rate onto the white dwarf \citep{lasota2001}. If the instability develops in the outer regions of the accretion disk (outside-in outbursts), the rise to maximum brightness is rapid. However, the light curves of inside-out outbursts (which start in the inner disk regions) are more symmetric, with a relatively slow rise to the peak \citep{lasota2001}. Dwarf novae in outburst reach optical absolute magnitudes from $+6$ to $0$, depending on their orbital period (longer orbital periods mean larger accretion disks, and hence, they become brighter; \citealt{patterson2011}). The X-ray luminosities of dwarf novae in outbursts are typically $10^{29}\!-\!10^{32}$\,erg\,s$^{-1}$, about six orders of magnitude lower than those of classical novae \citep{schwope2024,rodriguez2024}.

Therefore, the discovery of supersoft X-ray emission from the optical transient ASASSN-16oh \citep{maccarone2019} was a surprise. The X-ray emission was consistent with that of a $\sim$900,000\,K blackbody and a luminosity of about $6.7 \times 10^{36}$\,erg\,s$^{-1}$, similar to that seen from persistent supersoft X-ray sources and classical novae. However, the optical properties of the transient (light-curve shape, its amplitude, peak absolute magnitude, and optical spectra) did not match those of classical novae at all. 

ASASSN-16oh was first detected in 2016 December by the All-Sky Automated Survey for Supernovae (ASASSN; \citealt{shappee2014}) as a $V=16.9$ transient in the Small Magellanic Cloud (SMC). Its archival light curve (upper panel of Figure~\ref{fig:light_curves}) collected by the Optical Gravitational Lensing Experiment (OGLE; \citealt{udalski2015}) revealed that the optical outburst commenced at least four months before the ASASSN discovery. The object exhibited irregular variability before (and after) the outburst, with a mean brightness of about $I=20.3$ and $V=21.1$. It peaked at $I \approx 16.6$, which corresponds to $M_I=-2.5 \pm 0.1$, assuming the SMC distance modulus of $\mu = 18.977 \pm 0.016$ \citep{graczyk2020} and reddening of $E(V-I)=0.059^{+0.052}_{-0.045}$ \citep{skowron2021}. The full optical light curve does not match those of classical novae, which normally rise to maximum within hours to a few days (not months). Additionally, the optical spectra revealed narrow ($\mathrm{FWHM}=164$\,km\,s$^{-1}$) Balmer and He\,\textsc{ii} emission lines \citep{maccarone2019}, unlike the broad features (several hundred to a few thousand km\,s$^{-1}$) usually observed in classical novae.

The unusual outburst light curve, narrow optical emission lines, and X-ray spectrum led \citet{maccarone2019} to propose that the X-rays come from a spreading layer---a belt around the white dwarf's equator near the inner disk edge in which a large fraction of the total accretion energy is emitted. However, this interpretation remains controversial, with alternative models being based on a nonejecting nova outburst \citep{hillman2019,kato2020}.

\citet{hillman2019} proposed that both the optical flux and X-ray emission in ASASSN-16oh originated from a hot white dwarf that underwent a non-mass-ejecting thermonuclear runaway. They argued that such nonejecting nova events are possible if the mass accretion rate onto the white dwarf is sufficiently high (at least a few times $10^{-7}\,M_{\odot}\,\mathrm{yr}^{-1}$) and claimed that the light curve of ASASSN-16oh is best modeled by a $1.1\,M_{\odot}$ white dwarf accreting at $3.5\!-\!5\times 10^{-7}\,M_{\odot}\,\mathrm{yr}^{-1}$. Nonetheless, their model predicted an asymmetric outburst (with a rapid rise to the peak followed by a slower decline), in stark contrast to observations. Furthermore \citet{kato2020} later criticized this model, pointing out that Hillman et al.~had overestimated the optical brightness by 6\,mag.

On the other hand, \citet{kato2020} proposed that the outburst of ASASSN-16oh was triggered by a sudden mass accretion caused by disk instability. In this model, the optical flux originates from the irradiated accretion disk, and X-rays are coming from the hot hydrogen-burning white dwarf surface. They found that the X-ray light curve of ASASSN-16oh is best explained by a massive ($1.32\,M_{\odot}$) white dwarf model.

Given the unusual and unexplained optical and X-ray properties of ASASSN-16oh, we decided to investigate whether this star is one of a kind or representative of a larger, as yet undiscovered, group. We thus searched for objects with outburst properties similar to those seen in ASASSN-16oh among the 20-year-long light curves of 76~million stars observed toward the Magellanic Clouds by the OGLE survey \citep{udalski2015}. 

\section{OGLE Photometric Data}

The OGLE photometric data have been collected using the 1.3 m Warsaw Telescope located at Las Campanas Observatory, Chile, during the OGLE-III (2001--2009; \citealt{udalski2003}) and OGLE-IV (2010--2024; \citealt{udalski2015}) phases of the project. Different instruments were used in different phases of the survey. The \mbox{OGLE-III} camera consisted of eight $2048\!\times\!4096$ CCD detectors, with a plate scale of $0.26$\,arcsec\,pixel$^{-1}$. The OGLE-IV camera had the same pixel scale but comprised thirty-two $2048\!\times\!4102$ detectors, providing a field of view of 1.4\,deg$^2$. The typical exposure time was 150\,s.  The photometric data were extracted using a custom version of the difference image analysis algorithm \citep{tomaney1996,alard1998} implemented by \citet{wozniak2000}. The OGLE-III and OGLE-IV data have been reduced using the same reference image \citep{mroz2024}, enabling us to obtain homogeneous, long-term light curves of all analyzed objects. Photometry was calibrated to the standard $I$- and $V$-band systems. See \citet{udalski2003} and \citet{udalski2015} for a detailed description of the observing setup and data calibration. All data presented in this paper are available to the astronomical community from \url{https://ftp.astrouw.edu.pl/ogle/ogle4/millinovae/}.

\section{Results}

We started with a list of 72,303 objects (54,996 in the Large Magellanic Cloud, LMC, and 17,307 in the SMC) whose light curves contain at least five consecutive data points magnified with respect to the remaining light curve (see \citealt{mroz2024} for more details about the selection of outbursting events). We required that the light curve contain at least one outburst with an amplitude larger than 1\,mag and duration between 10 and 600\,days. Subsequently, we cross-matched selected objects with the \textit{Gaia} Data Release 3 archive \citep{gaia2023} and removed artifacts due to high-proper-motion stars (those with proper motions $>$10\,mas\,yr$^{-1}$), which were the largest source of contamination. The light curves of the remaining objects (7412 in the LMC and 2689 in the SMC) were visually inspected, and we selected the initial sample of objects with triangle-shaped, symmetrical outbursts similar to those of ASASSN-16oh.

We vetted all selected objects using additional data, including multicolor photometry, \textit{Gaia} parallaxes and proper motions, and sky images. We removed objects that could be classified as classical novae, supernovae, and active galactic nuclei, leaving us with a list of 29 objects (22 in the LMC and seven in the SMC). Example light curves are presented in Figure~\ref{fig:light_curves} (see also Figures~\ref{fig:lc1} and~\ref{fig:lc2} in Appendix~\ref{sec:app_lc}). All objects show long-duration (weeks to months), symmetrical (triangle-shaped) outbursts with amplitudes in $I$ ranging from 1.0 to 3.7\,mag. Some outbursts are deceivingly symmetric, which led to them being classified as gravitational microlensing events (OGLE-mNOVA-12 = MACHO-LMC-7 and OGLE-mNOVA-13 = MACHO-LMC-23, \citealt{alcock2000}; OGLE-mNOVA-25 = OGLE-LMC-02, \citealt{wyrzykowski2009}) although such events should be achromatic and nonrepeating if they really are due to microlensing. The list of all our selected objects is presented in Table~\ref{tab:list}.

Given the unique spectral and X-ray characteristics of ASASSN-16oh, we also decided to start the near-real-time monitoring of these selected objects in 2023 September, with the goal of triggering follow-up observations should any enter an outburst state.

\clearpage

\movetabledown=2.5in
\begin{rotatetable*}
\begin{deluxetable*}{ccccccccrccl}
\tablecaption{List of Detected Objects\label{tab:list}}
\tabletypesize{\scriptsize}
\tablehead{\colhead{Name} & \colhead{R.A.} & \colhead{Decl.} & \colhead{$I_{\rm q}$} & \colhead{$(V-I)_{\rm q}$} & \colhead{$I_{\rm max}$} & \colhead{$(V-I)_{\rm max}$} &  \colhead{$\Delta I$} & \colhead{$\Delta T$} & \colhead{$\tau_{\rm d}$} & \colhead{$E(V-I)$} & \colhead{Comments} \\ 
\colhead{} & \colhead{} & \colhead{} & \colhead{(mag)} & \colhead{(mag)} & \colhead{(mag)} & \colhead{(mag)} &  \colhead{(mag)} & \colhead{(day)} & \colhead{(day\,mag$^{-1}$)} & \colhead{(mag)} & \colhead{}}
\startdata
OGLE-mNOVA-01 & \ra{01}{57}{43}{64} & \dec{-73}{37}{32}{5} & $20.273 \pm 0.016$ & $0.784 \pm 0.056$ & $16.595 \pm 0.012$ & $0.198 \pm 0.008$ & $3.7$ & $>200$ & $33$ & 0.059 & ASASSN-16oh \\
OGLE-mNOVA-02 & \ra{00}{20}{40}{41} & \dec{-75}{11}{56}{5} & $20.981 \pm 0.017$ & $1.088 \pm 0.058$ & $19.216 \pm 0.050$ & $\dots$           & $1.8$ & $>70$ & $\dots$ &  0.042 & \dots\\
OGLE-mNOVA-03 & \ra{00}{26}{10}{69} & \dec{-73}{34}{18}{2} & $21.063 \pm 0.040$ & $0.502 \pm 0.186$ & $19.956 \pm 0.055$ & $0.501 \pm 0.055$ & $1.1$ & $55$ & $33$ & 0.028 & \dots\\
OGLE-mNOVA-04 & \ra{00}{34}{30}{23} & \dec{-74}{05}{40}{3} & $21.008 \pm 0.025$ & $0.611 \pm 0.239$ & $19.706 \pm 0.064$ & $\dots$           & $1.3$ & $83$ & $26$ & 0.046 & \dots\\
OGLE-mNOVA-05 & \ra{00}{50}{08}{58} & \dec{-69}{46}{33}{8} & $20.745 \pm 0.017$ & $0.894 \pm 0.068$ & $18.885 \pm 0.022$ & $0.420 \pm 0.017$ & $1.9$ & $74$ & $39$ & 0.016 & candidate\\
OGLE-mNOVA-06 & \ra{00}{51}{18}{58} & \dec{-68}{54}{34}{7} & $20.916 \pm 0.017$ & $1.051 \pm 0.073$ & $19.268 \pm 0.029$ & $0.354 \pm 0.033$ & $1.6$ & $102$ & $20$ & 0.011 & \dots\\
OGLE-mNOVA-07 & \ra{00}{52}{45}{30} & \dec{-72}{20}{07}{5} & $18.276 \pm 0.010$ & $0.574 \pm 0.015$ & $17.301 \pm 0.012$ & $0.545 \pm 0.008$ & $1.0$ & $>235$  & $213$ & 0.068 & \dots\\
OGLE-mNOVA-08 & \ra{04}{51}{40}{68} & \dec{-68}{25}{14}{5} & $20.197 \pm 0.013$ & $0.987 \pm 0.019$ & $18.255 \pm 0.024$ & $\dots$           & $1.9$ & $78$  & $27$ & 0.128 & \dots\\
OGLE-mNOVA-09 & \ra{04}{51}{58}{14} & \dec{-68}{30}{35}{6} & $20.057 \pm 0.015$ & $1.428 \pm 0.064$ & $19.015 \pm 0.046$ & $1.267 \pm 0.028$ & $1.0$ & $115$ & $105$ & 0.122 & candidate \\
OGLE-mNOVA-10 & \ra{04}{56}{24}{20} & \dec{-68}{27}{31}{5} & $20.673 \pm 0.026$ & $0.789 \pm 0.110$ & $19.586 \pm 0.091$ & $0.771 \pm 0.128$ & $1.1$ & $80$ & $41$ &  0.119 & \dots\\
OGLE-mNOVA-11 & \ra{04}{59}{56}{68} & \dec{-67}{31}{48}{9} & $20.884 \pm 0.039$ & $1.244 \pm 0.129$ & $18.151 \pm 0.021$ & $0.420 \pm 0.023$ & $2.7$ & $142$ & $29$ & 0.100 & \dots \\
OGLE-mNOVA-12 & \ra{05}{04}{03}{38} & \dec{-69}{33}{17}{9} & $20.814 \pm 0.033$ & $0.776 \pm 0.171$ & $19.407 \pm 0.029$ & $0.570 \pm 0.014$ & $1.4$ & $95$ & $65$ & 0.083 & MACHO-LMC-7\\
OGLE-mNOVA-13 & \ra{05}{06}{17}{46} & \dec{-70}{58}{46}{8} & $20.096 \pm 0.011$ & $0.847 \pm 0.018$ & $18.909 \pm 0.048$ & $\dots$           & $1.2$ & $131$ & $30$ & 0.124 & MACHO-LMC-23\\
OGLE-mNOVA-14 & \ra{05}{10}{15}{41} & \dec{-70}{31}{43}{6} & $20.423 \pm 0.018$ & $0.372 \pm 0.061$ & $18.847 \pm 0.058$ & $0.156 \pm 0.091$ & $1.6$ & $95$ & $17$ &  0.092 & \dots \\
OGLE-mNOVA-15 & \ra{05}{12}{44}{80} & \dec{-69}{41}{28}{0} & $20.790 \pm 0.022$ & $0.886 \pm 0.078$ & $18.540 \pm 0.018$ & $0.478 \pm 0.015$ & $2.2$ & $>120$ & $34$ & 0.183 & candidate \\
OGLE-mNOVA-16 & \ra{05}{14}{22}{96} & \dec{-70}{56}{56}{1} & $20.545 \pm 0.019$ & $0.327 \pm 0.031$ & $19.202 \pm 0.045$ & $0.223 \pm 0.061$ & $1.3$ & $42$ & $24$ &  0.072 & \dots\\
OGLE-mNOVA-17 & \ra{05}{15}{05}{58} & \dec{-68}{31}{07}{2} & $20.461 \pm 0.015$ & $0.893 \pm 0.087$ & $19.312 \pm 0.065$ & $0.810 \pm 0.027$ & $1.1$ & $120$ & $49$ & 0.105 & \dots\\
OGLE-mNOVA-18 & \ra{05}{15}{17}{91} & \dec{-70}{36}{58}{6} & $20.366 \pm 0.011$ & $0.904 \pm 0.018$ & $18.570 \pm 0.037$ & $0.756 \pm 0.078$ & $1.8$ & $105$ & $58$ & 0.094 & \dots\\
OGLE-mNOVA-19 & \ra{05}{17}{12}{72} & \dec{-68}{49}{38}{4} & $21.179 \pm 0.031$ & $0.838 \pm 0.084$ & $19.775 \pm 0.037$ & $0.541 \pm 0.045$ & $1.4$ & $48$ & $21$ & 0.102 & \dots\\
OGLE-mNOVA-20 & \ra{05}{20}{05}{81} & \dec{-69}{38}{31}{0} & $19.648 \pm 0.011$ & $0.118 \pm 0.018$ & $18.259 \pm 0.020$ & $0.146 \pm 0.024$ & $1.4$ & $108$ & $35$ & 0.078 & \dots\\
OGLE-mNOVA-21 & \ra{05}{25}{58}{44} & \dec{-69}{34}{33}{8} & $19.886 \pm 0.011$ & $0.656 \pm 0.017$ & $18.492 \pm 0.025$ & $0.498 \pm 0.020$ & $1.4$ & $>110$ & $39$ & 0.078 & \dots\\
OGLE-mNOVA-22 & \ra{05}{26}{45}{21} & \dec{-70}{29}{45}{7} & $18.574 \pm 0.010$ & $0.580 \pm 0.014$ & $17.653 \pm 0.015$ & $\dots$           & $0.9$ & $>800$ & $\dots$ & 0.139 & \dots\\
OGLE-mNOVA-23 & \ra{05}{27}{48}{98} & \dec{-68}{15}{44}{6} & $21.167 \pm 0.047$ & $0.700 \pm 0.302$ & $20.006 \pm 0.051$ & $0.451 \pm 0.029$ & $1.2$ & $33$ & $14$ & 0.099 & \dots\\
OGLE-mNOVA-24 & \ra{05}{28}{25}{12} & \dec{-70}{20}{43}{8} & $21.032 \pm 0.048$ & $0.671 \pm 0.094$ & $18.661 \pm 0.036$ & $0.487 \pm 0.027$ & $2.4$ & $>280$ & $48$ & 0.082 & \dots\\
OGLE-mNOVA-25 & \ra{05}{30}{47}{88} & \dec{-69}{54}{33}{8} & $20.404 \pm 0.014$ & $0.428 \pm 0.022$ & $19.228 \pm 0.080$ & $0.520 \pm 0.059$ & $1.2$ & $81$ & $39$ & 0.060 & OGLE-LMC-02\\
OGLE-mNOVA-26 & \ra{05}{32}{10}{63} & \dec{-70}{22}{09}{5} & $20.764 \pm 0.027$ & $0.097 \pm 0.116$ & $19.366 \pm 0.057$ & $-0.023 \pm 0.025$ & $1.4$ & $43$ & $148$ & 0.124 & \dots\\
OGLE-mNOVA-27 & \ra{05}{37}{56}{29} & \dec{-68}{48}{51}{0} & $20.763 \pm 0.028$ & $1.010 \pm 0.025$ & $19.273 \pm 0.069$ & $\dots$           & $1.5$ & $40$ & $14$ & 0.263 & \dots\\
OGLE-mNOVA-28 & \ra{05}{52}{29}{30} & \dec{-71}{10}{29}{9} & $20.818 \pm 0.023$ & $1.079 \pm 0.227$ & $19.439 \pm 0.034$ & $0.537 \pm 0.055$ & $1.4$ & $85$ & $26$ & 0.141 & \dots\\
OGLE-mNOVA-29 & \ra{05}{53}{41}{54} & \dec{-70}{22}{23}{0} & $20.747 \pm 0.019$ & $0.410 \pm 0.027$ & $19.758 \pm 0.078$ & $0.655 \pm 0.088$ & $1.0$ & $>120$ & $\dots$ & 0.105 & \dots\\
\enddata
\tablecomments{The table provides equatorial coordinates (for the epoch J2000), mean $I$-band brightness and $V-I$ color in quiescence, mean $I$-band brightness and $V-I$ color in outburst, amplitude of the outburst in $I$-band $\Delta I$, duration of the highest-amplitude outburst $\Delta T$, mean decline rate $\tau_{\rm d}$, and color excess $E(V-I)$ \citep{skowron2021} toward the detected objects.}
\end{deluxetable*}
\end{rotatetable*}

\section{Outburst of OGLE-\lowercase{m}NOVA-11}

\begin{figure}[htbp]
\centering
\includegraphics[width=.5\textwidth]{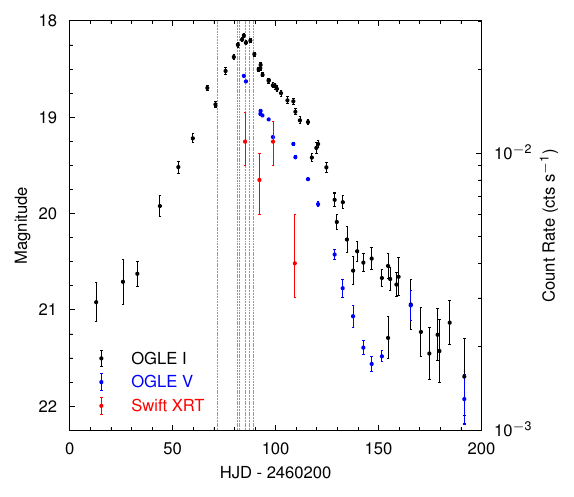}
\caption{Photometric and X-ray observations of the 2023/2024 outburst of OGLE-mNOVA-11. Dotted lines mark times of SALT spectroscopic observations.}
\label{fig:phot}
\end{figure}

The outburst of OGLE-mNOVA-11 started shortly afterwards, between 2023 October 15.3 and 26.3 and reached a maximum of $I=18.15 \pm 0.02$ and $V=18.57 \pm 0.01$ on 2023 December 6.1 (Figure~\ref{fig:phot}), an amplitude of almost 3 mag. This corresponds to $M_I=-0.5 \pm 0.1$ and $M_V=-0.2 \pm 0.1$, assuming the LMC distance modulus of $\mu=18.477 \pm 0.004$ \citep{pietrzynski2019} and reddening $E(V-I)=0.100^{+0.065}_{-0.044}$ \citep{skowron2021}. The star returned to quiescence (in $I$) in late 2024 February, about 120 days after the outburst onset. The outburst duration was shorter in $V$.

\begin{figure*}[htbp]
\centering
\includegraphics[width=.9\textwidth]{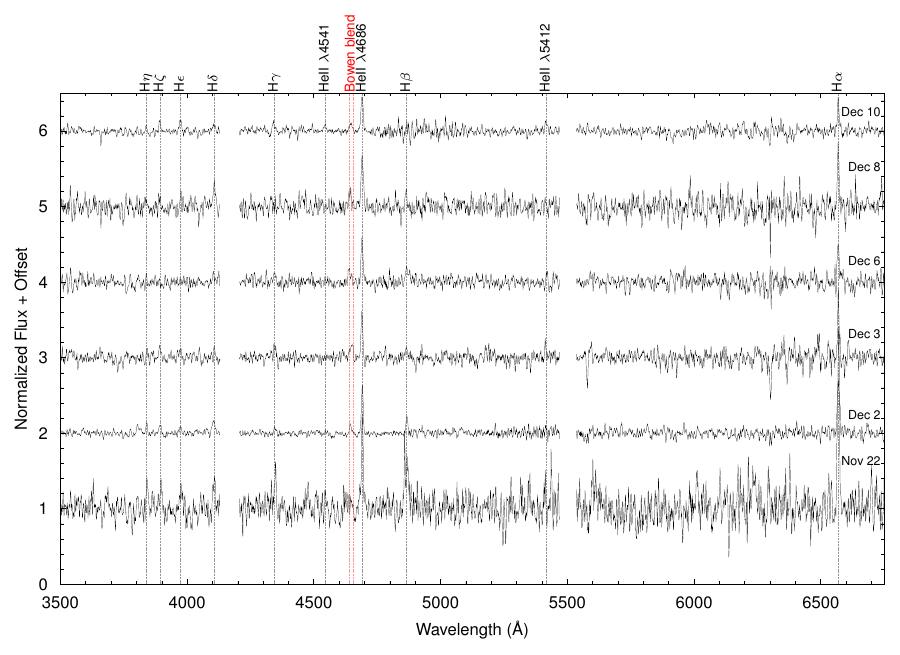}
\caption{SALT spectroscopic observations of the 2023/2024 outburst of OGLE-mNOVA-11 reveal narrow emission lines, including He\,\textsc{ii} (4686\,\AA) and Bowen blend.}
\label{fig:spectra}
\end{figure*}

We obtained a set of low-resolution spectra of OGLE-mNOVA-11 from 2023 November 22 to December 10 (Figure~\ref{fig:spectra}) with the Robert Stobie Spectrograph (\citealt{kobulnicky2003,burgh2003}) mounted on the Southern African Large Telescope (SALT), as part of the SALT Large Science Programme on Transients (Appendix~\ref{app:salt}). The spectra cover the range 3500--6700\,\AA{} and reveal narrow ($\mathrm{FWHM} = 247 \pm 28$\,km\,s$^{-1}$) emission lines (Balmer lines, He\,\textsc{ii} 4686\,\AA{}). The Bowen blend (a complex of C\,\textsc{iii} and N\,\textsc{iii} lines in the range $4640\!-\!4650$\,\AA) is also clearly detected from 2023 December 2 to 10. The emission lines are redshifted with a mean radial velocity of $278.3 \pm 4.5$\,km\,s$^{-1}$, close to that of the LMC systemic value ($262.2 \pm 3.4$\,km\,s$^{-1}$; \citealt{vdm2002}). Both the light-curve shape and optical spectra are very different from those of classical novae.

OGLE-mNOVA-11 was also observed by the Neil Gehrels Swift Observatory \citep{gehrels2004} five times from 2023 December 6 to 30 (Appendix~\ref{app:swift}).  Swift's X-Ray Telescope (XRT; \citealt{burrows2005}) detected a faint X-ray source at the transient's position, with a 0.3--10\,keV count rate declining from $0.011^{+0.003}_{-0.002}$ count\,s$^{-1}$ (December 6) to $0.004^{+0.002}_{-0.001}$ count\,s$^{-1}$ (December 30); see Figure~\ref{fig:phot}. All the data were combined to create a single X-ray spectrum, which can be fitted with a blackbody of $418{,}000 ^{+220{,}000}_{-290{,}000}$\,K absorbed by a column density $N_{\rm H} = 3.5^{+13.0}_{-3.0}\times 10^{21}$\,cm$^{-2}$. If we fix $N_{\rm H}$ to the LMC value ($1.3 \times 10^{21}$\,cm$^{-2}$), then the temperature is slightly better constrained to $607{,}000 ^{+160{,}000}_{-130{,}000}$\,K. Assuming this blackbody model, the observed L$_X$ (0.3--10\,keV) $\approx 0.9 \times 10^{35}$\,erg\,s$^{-1}$, or $3.6 \times 10^{35}$\,erg\,s$^{-1}$ when corrected for absorption, about an order of magnitude smaller than that of ASASSN-16oh, but still substantially higher than dwarf nova X-ray luminosities.

\section{Discussion and Conclusions}

Overall, the optical (light-curve shape, narrow emission lines, strong He\,\textsc{ii} emission) and X-ray properties of OGLE-mNOVA-11 match well with those of ASASSN-16oh. We thus propose that they, and probably the remaining objects described in this study, form a homogeneous group of transient supersoft X-ray sources. We dub them ``millinovae'' because their optical luminosities are roughly a thousand times fainter than those of ordinary classical novae.\footnote{They should not be confused with ``micronovae,'' a much lower-luminosity and much faster event seen in a number of Galactic magnetic cataclysmic variables \citep{scaringi2022}.}

Using ASASSN-16oh and OGLE-mNOVA-11 as prototypes, millinovae should fulfill the following criteria:
\begin{enumerate}
    \itemsep0em
    \item They exhibit symmetrical, triangle-shaped outbursts in the optical bands ($V$, $I$).
    \item Outbursts last from a month to several months, i.e., substantially longer than the typical outbursts of short-orbital-period dwarf novae.
    \item The peak absolute magnitude in the optical bands is between that of classical novae and short-orbital-period dwarf novae.
    \item The optical spectra show narrow (FWHM smaller than a few hundred km\,s$^{-1}$) emission lines of H and He\,\textsc{ii} (that is, no signatures of mass ejection are seen in the spectra).
    \item Soft X-ray emission (declining with decreasing optical luminosity) is seen during the outburst.
\end{enumerate}

Only two systems (ASASSN-16oh and OGLE-mNOVA-11) meet all these five criteria. The remaining objects, given the lack of dedicated spectroscopic and X-ray observations during outbursts, remain millinova candidates.

\begin{figure*}[htbp]
\centering
\includegraphics[width=0.7\textwidth]{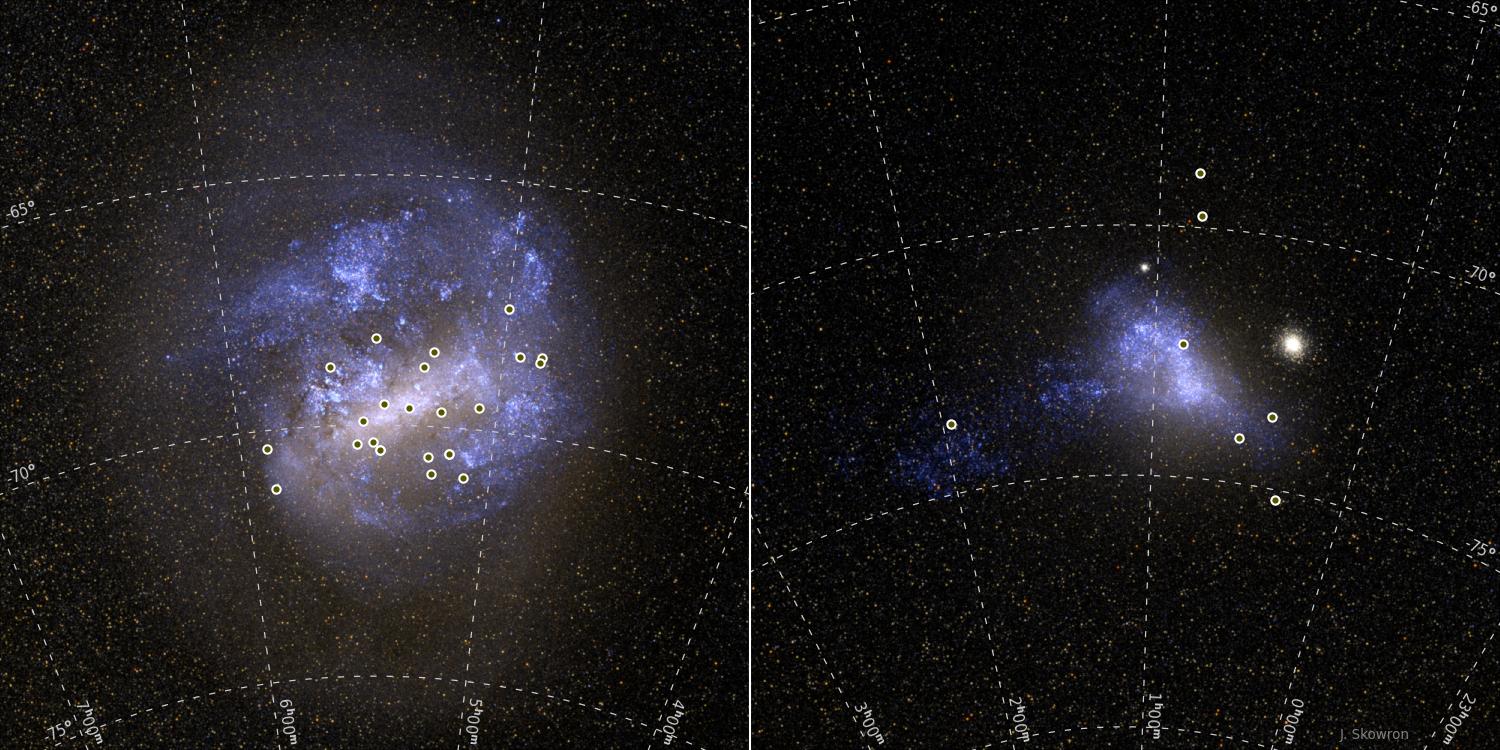}
\includegraphics[width=.8\textwidth]{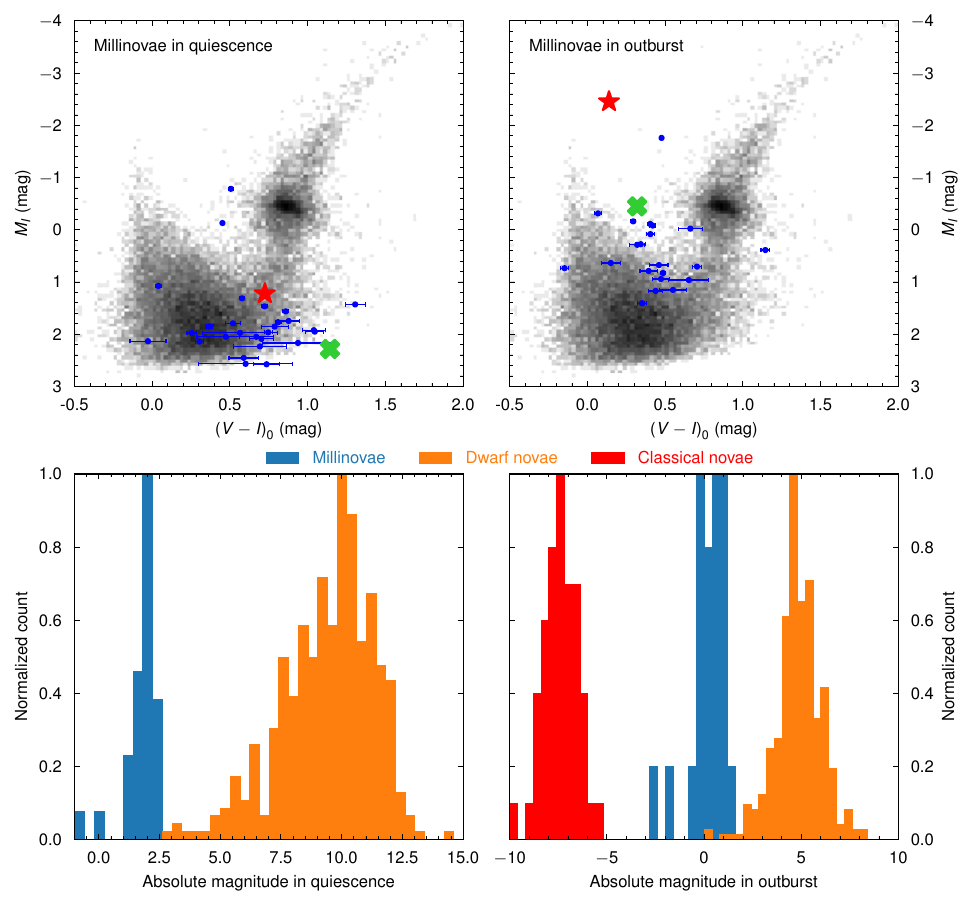}
\caption{Upper panels: on-sky view of the LMC (upper left) and SMC (upper right) with positions of millinovae marked. Middle panels: dereddened color--magnitude diagrams for stars in the LMC. Blue points mark $(V-I)_0$ and $M_I$ of millinovae in quiescence (middle left) and in outburst (middle right). The prototype, ASASSN-16oh, is marked with a red asterisk, and OGLE-mNOVA-11 is marked with a green cross. Lower panels: histograms of $M_I$ of millinovae, classical novae, and dwarf novae in quiescence (lower left) and in outburst (lower right). Data were taken from \citet{ritter_kolb_2003}, \citet{pietsch2010}, and \citet{gaia2023}. The background images of the LMC and SMC were generated with \textsc{bsrender} written by Kevin Loch, using the ESA/\textit{Gaia} database.}
\label{fig:map}
\end{figure*}

Their sky location (upper panels of Figure~\ref{fig:map}) and the fact that most of the selected stars occupy a relatively narrow region in the color--magnitude diagram (middle left panel of Figure~\ref{fig:map}) indicate that millinovae are all in the Magellanic Clouds. We used the long-term astrometric time series of OGLE observations to measure proper motions of selected objects (Appendix~\ref{app:pm}), which are consistent (with one exception) with those of stars located in the Magellanic Clouds \citep{helmi2018}. The radial velocities of ASASSN-16oh \citep{maccarone2019} and OGLE-mNOVA-11 (Appendix~\ref{app:salt}) are consistent with the systemic velocities of the SMC and LMC, respectively. However, we cannot rule out that some objects are located in the foreground Milky Way disk or in background galaxies although we consider this very unlikely.

The middle left panel of Figure~\ref{fig:map} shows that most objects in quiescence occupy the same region of the color--magnitude diagram, $0.5 \leq (V-I)_0 \leq 1.0$, $1.5 \leq M_I \leq 2.5$, which indicates that the accretion disk dominates their quiescent luminosity. Such a range is consistent with the absolute magnitudes of dwarf novae in the Magellanic Clouds \citep{shara2003}, indicating that they are essentially nova-like disks. In outburst (middle right panel of Figure~\ref{fig:map}), objects move toward bluer colors ($0.0 \leq (V-I)_0 \leq 0.5$) and higher luminosities ($-0.5 < M_I < 1.0$), as expected given the strong irradiation of the disk by the supersoft component.

The mean absolute magnitudes of millinovae in quiescence are $M_I = 1.72 \pm 0.70$ and $M_V = 2.37 \pm 0.80$. In outburst, the absolute magnitudes are $M_I = 0.18 \pm 0.83$ and $M_V = 0.62 \pm 0.97$ and are about a thousand times fainter than classical novae (lower right panel of Figure~\ref{fig:map}). Millinovae are also much brighter than most known dwarf novae (lower panels of Figure~\ref{fig:map}).

\begin{figure}[htbp]
\centering
\includegraphics[width=.5\textwidth]{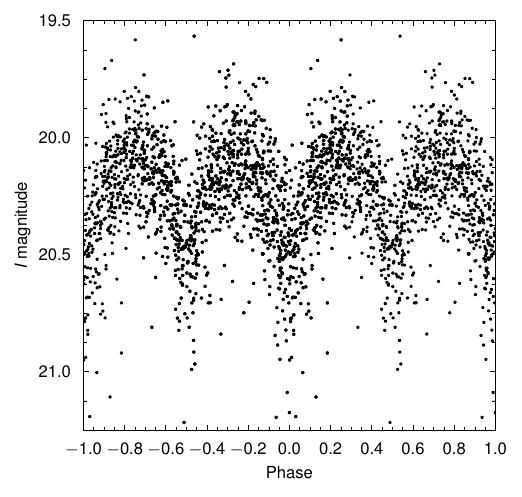}
\caption{Quiescent light curve of OGLE-mNOVA-08 folded with the orbital period $P_{\rm orb}=4.830893$\,days.}
\label{fig:eclipsing}
\end{figure}

The spectroscopic properties of ASASSN-16oh and OGLE-mNOVA-11 and the $(V-I)$ color evolution of OGLE-mNOVA-11 indicate that their optical outbursts likely have a disk origin. Because dwarf novae follow a well-known relation between the orbital period and absolute magnitude (both in quiescence and outburst; \citealt{patterson2011}), relatively high absolute magnitudes of millinovae indicate that their orbital periods should be on the order of a few days. Indeed, one of the analyzed objects (OGLE-mNOVA-08) shows clear eclipsing variability with a $4.830893(47)$ day period (Figure~\ref{fig:eclipsing}). (We searched for possible periodic brightness variations in quiescence and during outbursts in all 29 objects but detected a statistically significant signal only in the case of OGLE-mNOVA-08.)

\begin{figure}[htbp]
\centering
\includegraphics[width=.5\textwidth]{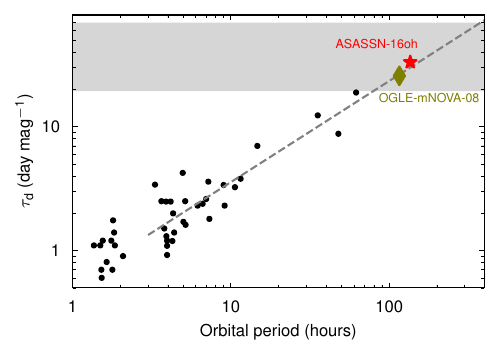}
\caption{Orbital period ($P_{\rm orb}$)--decline rate ($\tau_{\rm d}$) relation for dwarf nova outbursts, using data from \citet{simon2021}, and extended to include the range of decline rates (gray area) observed in millinovae ($20 \lesssim \tau_{\rm d} \lesssim 70$\,day\,mag$^{-1}$). The red asterisk and green diamond mark the locations of ASASN-16oh \citep{rajoelimanana2017} and OGLE-mNOVA-08, respectively.}
\label{fig:bailey}
\end{figure}

Furthermore, the rate of decline from the outburst of dwarf novae, ${\tau}_{\rm d}$, is also known to be proportional to $P_{\rm orb}$ (the ``Bailey relation''; \citealt{bailey1975}; see Figure~\ref{fig:bailey}), as expected, given that longer periods imply larger disks. We find that most millinovae have decline rates in the range $20<\tau_{\rm d}<70$\,day\,mag$^{-1}$, which corresponds to $P_{\rm orb}$ from 3 to 15 days, assuming that the Bailey relation can be extrapolated to such long periods.

The ``spreading layer'' \citep{kippenhahn1978,inogamov1999,piro2004,maccarone2019} model can naturally produce the soft X-ray temperatures that have been seen in many dwarf nova outbursts, albeit in shorter-period systems such as U\,Gem and SS\,Cyg (see, e.g., \citealt {wheatley2003}).  However, the long orbital periods of millinovae imply the existence of large disks and hence likely large mass-transfer rates (e.g., \citealt{echevarria1994}) capable of producing higher X-ray luminosities.  Nevertheless, there are difficulties with this model that have been noted (e.g., \citealt{kato2020}) in that there are high mass-transfer-rate recurrent novae in quiescence (e.g. U\,Sco, RS\,Oph) where such a soft component would have been expected but has never been seen.  

The most obvious solution to explaining the properties of millinovae is then to find a way to trigger thermonuclear burning on the white dwarf surface without producing a nova flash.  One such attempt to do this is the nonejecting thermonuclear runaway event of \citet{hillman2019}, but the resulting light curves of their ASASSN-16oh model are completely different to the smooth rise and fall that are observed, not to mention the huge color variation they predict, which is definitely excluded by the data.  

An alternative route to such a trigger has been sought by \citet{kato2020}, who build on the already known, ongoing high mass-transfer rate in these systems (their brightness and variability properties are consistent with nova-like disks in the Magellanic Clouds; \citealt{shara2003}), thereby keeping the disk (and white dwarf) in a hot, active state.  Furthermore, they require a massive white dwarf, close to 1.2\,$M_\odot$ in order to initiate steady burning without a strong wind or shell being ejected.  While we do not yet know how many of these long symmetric outburst systems produce detectable supersoft X-rays, if they were all to do so, then this would have important implications for the theory of nonejecting supersoft X-ray sources, as it would constrain the parameters of systems capable of behaving in this way.  Furthermore, it would provide a route for allowing the white dwarf to continue to grow in mass and hence become a potential Type Ia supernova progenitor.  

We recognize, however, that an added complication to the \citet{kato2020} models is that they produce close to Eddington-limited events, whereas some of our millinovae clearly do not reach such a level in X-rays (Appendix~\ref{app:erosita}). There may be severe geometrical constraints on the fluxes observed, or the nuclear burning might not cover the entire white dwarf surface. 
Moreover, some of these models (see also \citealt{kato2022}) predict the highest, close-to-Eddington flux to be in the far UV range (rather than soft X-rays), which can be tested with dedicated follow-up observations of future millinova outbursts.
Nevertheless, we believe that this new group of millinovae, all likely representing long-period, high mass-transfer-rate cataclysmic variables, opens an important new route for study, with the added benefit of being a well-constrained population in the Magellanic Clouds.

\section*{Acknowledgments}
We thank all the OGLE observers for their contribution to the collection of the photometric data over the decades. This research was funded in part by National Science Centre, Poland, grant OPUS 2021/41/B/ST9/00252 awarded to P.M. K.L.P. acknowledges funding from the UK Space Agency. A part of this work is based on observations made with the Southern African Large Telescope (SALT), with the Large Science Programme on transients 2021-2-LSP-001 (PI: D.A.H.B.). Polish participation in SALT was funded by grant No. MEiN 2021/WK/01.

\vspace{5mm}
\facilities{OGLE, Swift(XRT and UVOT), SALT(RSS), eROSITA}

\newpage
\clearpage

\appendix
\restartappendixnumbering

\section{Light Curves}
\label{sec:app_lc}

\begin{figure*}[htbp]
\centering
\includegraphics[width=\textwidth]{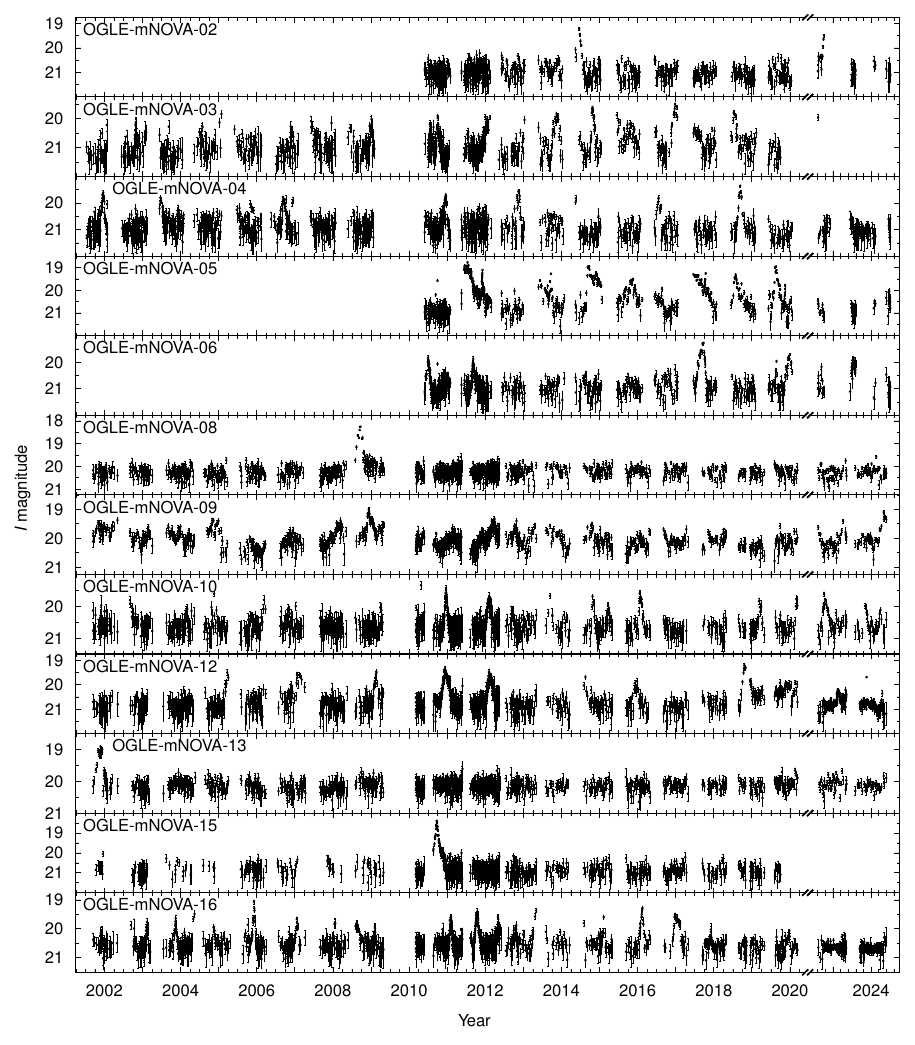}
\caption{Example light curves of millinovae (continuation).}
\label{fig:lc1}
\end{figure*}

\begin{figure*}[htbp]
\centering
\includegraphics[width=\textwidth]{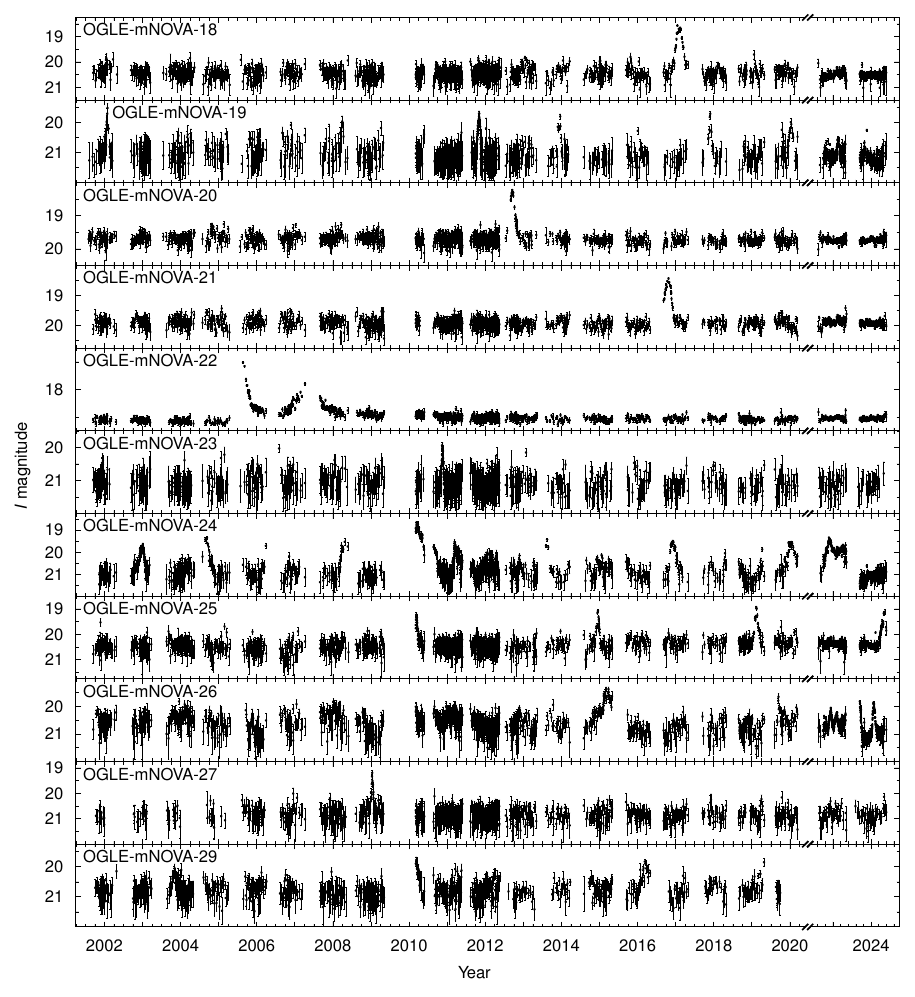}
\caption{Example light curves of millinovae (continuation).}
\label{fig:lc2}
\end{figure*}

Figures~\ref{fig:lc1} and~\ref{fig:lc2} show the light curves of the remaining stars exhibiting long-duration symmetrical outbursts, which we have classified as ``millinovae.''

\section{SALT Observations of OGLE-\lowercase{m}NOVA-11}
\label{app:salt}

Optical spectra of OGLE-mNOVA-11 were obtained with the Robert Stobie Spectrograph \citep{kobulnicky2003,burgh2003} mounted on SALT (\citealt{buckley2006}) in long-slit mode. Two 1200\,s exposure spectra were taken each night on 2023 November 22 and December 2, 3, 6, 8, and 10, using the PG0700 grating at an angle of $3^{\circ}$ with a slit width of $1.5''$, giving a resolving power of $460 < R < 800$. The data were prereduced using the PySALT package \citep{crawford2010}, which includes bias subtraction, gain and cross-talk corrections, and mosaicking. An Argon lamp was utilized for wavelength calibration, and the spectrophotometric standard star, EG~21, was used for flux calibration, both performed with IRAF. The spectra were stacked so as to obtain one average spectrum per night.

The software package, Fityk\footnote{https://fityk.nieto.pl/} \citep{wojdyr2010}, was used for spectral analysis. The SALT spectra were normalized before the emission lines were modeled using the method of \citet{Shafter1983} in order to determine the central wavelengths of the He\,\textsc{ii} (4686\,\AA), $\mathrm{H}\beta$, and $\mathrm{H}\alpha$ emission lines. This entailed fitting a Gaussian function through 95\% of the line wings. In Fityk, the errors of the fitted parameters were obtained for a confidence level of $1\sigma$ (68.27\%). The measured central wavelengths were then utilized to calculate the barycentric-corrected radial velocities.

Table~\ref{tab:salt} presents the barycentric radial velocities of He\,\textsc{ii} (4686\,\AA), H$\beta$, and H$\alpha$. Because of the larger scatter in our H$\beta$ velocities (due to its lower-significance detections), we also calculated the weighted mean radial velocity from He\,\textsc{ii} and H$\alpha$, obtaining values that ranged from $232.1$ to $311.3$\,km\,s$^{-1}$, with a mean of $278.3 \pm 4.5$\,km\,s$^{-1}$, which is very close to the established LMC line-of-sight velocity ($262.2 \pm 3.4$\,km\,s$^{-1}$; \citealt{vdm2002}). The number of radial velocity measurements prevents us from determining the orbital period unequivocally, but it is likely to be shorter than 4\,days. If we restrict orbital periods to be longer than 1\,day and assume a circular orbit, the best-fit periods are 2.42, 1.68, 2.08, and 1.89\,days.

\begin{table*}[htbp]
\center
\caption{SALT Radial Velocities of OGLE-mNOVA-11}
\label{tab:salt}
\begin{tabular}{lccccc}
\hline \hline
\multicolumn{1}{c}{Time} & BJD & He\,\textsc{ii} & H$\beta$ & H$\alpha$ & Mean RV\\
 & & (km\,s$^{-1}$) & (km\,s$^{-1}$) & (km\,s$^{-1}$) & (km\,s$^{-1}$) \\
\hline
2023 Nov 22.93 & 2460271.43 & $313.7 \pm 25.7$ & $199.9 \pm 132.0$ & $310.9 \pm 11.7$ & $311.3 \pm  10.6$ \\
2023 Dec 2.88 & 2460281.38 & $308.5 \pm  8.6$ & $329.4 \pm  25.6$ & $306.7 \pm 18.2$ & $308.2 \pm   7.8$ \\
2023 Dec 3.84 & 2460282.34 & $203.0 \pm 15.3$ & $203.0 \pm  99.5$ & $279.8 \pm 18.4$ & $234.3 \pm  11.8$ \\
2023 Dec 6.84 & 2460285.34 & $224.1 \pm 19.1$ & $475.5 \pm  95.5$ & $252.8 \pm 30.9$ & $232.1 \pm  16.3$ \\
2023 Dec 8.83 & 2460287.33 & $284.2 \pm 22.3$ & $143.1 \pm  59.5$ & $241.8 \pm 22.7$ & $263.4 \pm  15.9$ \\
2023 Dec 10.83 & 2460289.33 & $245.2 \pm 13.4$ & $417.6 \pm 124.2$ & $257.4 \pm 19.7$ & $249.0 \pm  11.1$ \\
\hline
\end{tabular}
\end{table*}

\restartappendixnumbering

\section{Swift Observations of OGLE-\lowercase{m}NOVA-11}
\label{app:swift}

\subsection{Swift X-ray Data}

The Neil Gehrels Swift Observatory \citep{gehrels2004} performed five Target of Opportunity observations of OGLE-mNOVA-11 (Target ID 16408) between 2023 December~6 and December~30, collecting $\approx 1.5$\,ks during each pointing, with the exception of December~17, when only $\approx0.5$\,ks was obtained. The XRT (\citealt{burrows2005}) was operated in Photon Counting mode, and the data were processed using HEASoft 6.32, together with the most up-to-date calibration files available. Given the faintness of the source, a small extraction region of 10 pixels ($23.6''$) radius was used for the source, and a larger, 60 pixel ($141.4''$) radius circle, offset from, but close to, the source was used to estimate the background contribution.

In observations where the source was detected at the $3\sigma$ level, the 0.3--10\,keV count rate (corrected for all point-spread function losses) is given in Table~\ref{tab:xrt}. In the shorter exposure on December~17, the source was not detected, and a $3\sigma$ upper limit is given instead.

There is no significant change in the hardness ratio during these observations though it should be noted that we are dealing with a limited number of source counts ($<50$ in total across all five observations, the vast majority below 1\,keV). Therefore, all the data were combined to create a single spectrum. Cash statistics \citep{cash1979} were used to fit this spectrum with a combination of an optically thick blackbody for the soft emission and an optically thin thermal plasma to account for the few higher-energy photons, both absorbed by the same hydrogen column. While the optically thin plasma can only be constrained to have a temperature of $>$~4.2\,keV ($> 4.9 \times 10^7$\,K), the best-fitting blackbody temperature is 36$^{+19}_{-25}$\,eV (418,000$^{+220,000}_{-290,000}$\,K), absorbed by a column density of $N_{\rm H} = (3.5^{+13.0}_{-3.0}) \times 10^{21}$\,cm$^{-2}$. This fit results in a C-stat value of 19.9 for 25 degrees of freedom.

If $N_{\rm H}$ is fixed to the LMC value of $1.3 \times 10^{21}$\,cm$^{-2}$ \citep{hi4pi2016}, the blackbody temperature becomes slightly better constrained, with $kT = 52^{+14}_{-11}$\,eV ($T =$ 607,000$^{+160,000}_{-130,000}$\,K), for C-stat/degrees of freedom = 21.2/26. This fit leads to an L$_X$ (0.3--10 keV) of $9.0 \times 10^{34}$\,erg\,s$^{-1}$, or $3.6 \times 10^{35}$\,erg\,s$^{-1}$ unabsorbed (assuming a distance of 49.59 kpc; \citealt{pietrzynski2019}).

\begin{table*}[htbp]
\center
\caption{Swift XRT Observations of OGLE-mNOVA-11}
\label{tab:xrt}
\begin{tabular}{cccc}
\hline \hline
Obs. ID & Start Time & Exp. Time & Count Rate \\
& (UT) & (s) & ($10^{-3}$ s$^{-1}$) \\
\hline
00016408001 & 2023-12-06 16:04:57 & 1594 & $11^{+3}_{-2}$ \\
00016408002 & 2023-12-13 09:21:57 & 1499 & $8 \pm 2$      \\
00016408003 & 2023-12-17 07:10:56 & 479  & $<20$            \\
00016408004 & 2023-12-20 04:27:56 & 1605 & $11 \pm 2$     \\
00016408005 & 2023-12-30 16:46:56 & 1529 & $4^{+2}_{-1}$  \\
\hline
\end{tabular}
\end{table*}

\subsection{Swift UV data}

OGLE-mNOVA-11 was also observed by the UltraViolet and Optical Telescope (UVOT; \citealt{roming2005}) on board the Neil Gehrels Swift Observatory. Observations took place during a period when one of the three onboard gyroscopes malfunctioned, resulting in increased noise and affecting the image quality of UVOT data \citep{cenko2023}. Moreover, the nearest source detected in the UVOT images was slightly offset from the target position by $3.6\!-\!4.8''$ and was likely a blend of the target and a nearby ($5.4''$) bright ($V = 16.996$, $V-I=-0.097$) constant star. 
We measured the brightness of the blend using the \textsc{uvotsource} tool, which performs aperture photometry. We used an aperture with a radius of $15''$ centered on the position of OGLE-mNOVA-11.

We then estimated the brightness of the neighbor from the analysis of the UV--optical color--magnitude diagrams, assuming that the neighbor's color is consistent with colors of other main-sequence stars of similar magnitude. We, therefore, created UV--optical color--magnitude diagrams by cross-matching the UVOT catalogs with the OGLE photometric map. We then calculated the mean $UV-V$ color of main-sequence stars in the brightness range $|V - 16.996| < \Delta$, where $\Delta = 0.15$\,mag. We found
\begin{align*}
uvm2 - V &= -1.462 \pm 0.078,\\
uvw1 - V &= -1.482 \pm 0.064,\\
uvw2 - V &= -1.559 \pm 0.077.
\end{align*}
(We also tested $\Delta=0.1$ and $\Delta=0.2$ and obtained similar results.) That enabled us to estimate the brightness of the neighbor in the UV filters:
\begin{align*}
uvm2 &= 15.534 \pm 0.078,\\
uvw1 &= 15.514 \pm 0.064,\\
uvw2 &= 15.437 \pm 0.077,
\end{align*}
and estimate the UV brightness of the target by subtracting the brightness of the neighbor star. Results are reported in Table~\ref{tab:uvot}. All magnitudes are in the Vega system.

\begin{table*}[htbp]
\center
\caption{Swift UVOT Observations of OGLE-mNOVA-11}
\label{tab:uvot}
\begin{tabular}{ccccccc}
\hline \hline
Obs. ID & $uvw1$ & $uvm2$ & $uvw2$\\
\hline
00016408001 & \dots              & $15.838 \pm 0.046$ & \dots \\
00016408002 & $16.448 \pm 0.094$ & $16.042 \pm 0.078$ & $16.019 \pm 0.068$\\
00016408003 & \dots              & $15.843 \pm 0.068$ & \dots \\
00016408004 & \dots              & $16.166 \pm 0.059$ & \dots \\
00016408005 & \dots              & $16.197 \pm 0.074$ & \dots \\
\hline
\end{tabular}
\end{table*}

\restartappendixnumbering

\section{OGLE Proper Motions}
\label{app:pm}

We calculated the proper motions of detected stars using the OGLE time-series astrometric data (OGLE-Uranus; A.~Udalski et al. 2024, in preparation). In short, we measured each star's position $(x,y)$ in individual OGLE images collected between 2010 and 2024 (in some cases, 2010--2020 if later data were unavailable). We then cross-matched the positions of bright stars with those calculated using the \textit{Gaia} Data Release 3 (DR3) data \citep{gaia2016,gaia2023}, taking into account their proper motion and heliocentric parallax. That allowed us to find the transformation between the positions $(x,y)$ in individual OGLE images to the equatorial coordinates in the \textit{Gaia} reference frame \citep{lindegren2021}. The proper motions in right ascension $\mu_{\alpha}$ and declination $\mu_{\delta}$ for all detected stars are reported in Table~\ref{tab:pm}. Because most of our targets are very faint ($I\approx20\!-\!21$), the proper motion uncertainties are large (the median error bar is 2\,mas\,yr$^{-1}$). Nonetheless, in all cases (but one), the proper motions are consistent within the quoted errors with the Magellanic Clouds' proper motions \citep{helmi2018}. The only exception is OGLE-mNOVA-15 with a total proper motion of $23.5 \pm 3.8$\,mas\,yr$^{-1}$, making it most likely to be a foreground object, and we therefore treat it as a candidate millinova.

\begin{deluxetable*}{lcccc}
\tablecaption{Millinovae Proper Motions\label{tab:pm}}
\tabletypesize{\footnotesize}
\tablehead{\colhead{Object} & \colhead{$\mu_{\alpha}$ (OGLE)} & \colhead{$\mu_{\delta}$ (OGLE)} & \colhead{$\mu_{\alpha}$ (\textit{Gaia})} & \colhead{$\mu_{\delta}$ (\textit{Gaia})}\\
\colhead{} & \colhead{(mas\,yr$^{-1}$)} & \colhead{(mas\,yr$^{-1}$)} & \colhead{(mas\,yr$^{-1}$)} & \colhead{(mas\,yr$^{-1}$)}}
\startdata
OGLE-mNOVA-01 & $ 1.66 \pm  2.03$ & $ 0.14 \pm  2.39$ & \dots & \dots \\
OGLE-mNOVA-02 & $ 1.31 \pm  2.35$ & $-2.50 \pm  2.31$ & \dots & \dots \\
OGLE-mNOVA-03 & $ 1.60 \pm  3.22$ & $-4.03 \pm  2.68$ & \dots & \dots \\
OGLE-mNOVA-04 & $ 1.25 \pm  2.78$ & $-3.15 \pm  2.53$ & \dots & \dots \\
OGLE-mNOVA-05 & $-0.37 \pm  1.44$ & $-0.44 \pm  1.49$ & $0.55 \pm 0.51$ & $-1.19 \pm 0.65$\\
OGLE-mNOVA-06 & $-0.67 \pm  2.04$ & $ 0.29 \pm  2.16$ & \dots & \dots \\
OGLE-mNOVA-07 & $ 0.74 \pm  0.09$ & $-1.39 \pm  0.10$ & $0.80 \pm 0.17$ & $-1.18 \pm 0.18$\\
OGLE-mNOVA-08 & $ 0.85 \pm  0.91$ & $ 0.84 \pm  0.92$ & \dots & \dots \\
OGLE-mNOVA-09 & $-1.26 \pm  1.16$ & $ 0.08 \pm  1.16$ & \dots & \dots \\
OGLE-mNOVA-10 & $ 2.13 \pm  1.26$ & $-0.13 \pm  1.51$ & \dots & \dots \\
OGLE-mNOVA-11 & $ 4.35 \pm  2.18$ & $ 4.08 \pm  2.37$ & \dots & \dots \\
OGLE-mNOVA-12 & $-3.00 \pm  2.51$ & $-7.29 \pm  2.64$ & \dots & \dots \\
OGLE-mNOVA-13 & $ 2.11 \pm  0.74$ & $-0.25 \pm  0.66$ & $2.53 \pm 2.28^{(a)}$ & $10.09 \pm 2.60^{(a)}$ \\
OGLE-mNOVA-14 & $ 1.96 \pm  2.56$ & $-2.29 \pm  1.99$ & \dots & \dots \\
OGLE-mNOVA-15 & $12.35 \pm  3.04$ & $-20.00 \pm 4.11$ & \dots & \dots \\
OGLE-mNOVA-16 & $ 3.32 \pm  1.37$ & $-9.88 \pm  3.42$ & \dots & \dots \\
OGLE-mNOVA-17 & $ 1.44 \pm  1.32$ & $ 2.71 \pm  1.87$ & \dots & \dots \\
OGLE-mNOVA-18 & $ 0.91 \pm  1.40$ & $ 0.38 \pm  1.67$ & \dots & \dots \\
OGLE-mNOVA-19 & $-4.35 \pm  3.39$ & $-1.48 \pm  2.37$ & \dots & \dots \\
OGLE-mNOVA-20 & $-0.34 \pm  1.67$ & $ 1.88 \pm  0.51$ & \dots & \dots \\
OGLE-mNOVA-21 & $ 0.19 \pm  0.61$ & $ 0.02 \pm  0.68$ & \dots & \dots \\
OGLE-mNOVA-22 & $ 1.82 \pm  0.09$ & $ 0.71 \pm  0.10$ & $2.24 \pm 0.19$ & $1.16 \pm 0.24$\\
OGLE-mNOVA-23 & $-1.93 \pm  4.22$ & $ 3.85 \pm  6.32$ & \dots & \dots \\
OGLE-mNOVA-24 & $ 6.43 \pm  4.11$ & $ 1.45 \pm  2.32$ & \dots & \dots \\
OGLE-mNOVA-25 & $-0.27 \pm  2.01$ & $-0.13 \pm  2.81$ & \dots & \dots \\
OGLE-mNOVA-26 & $ 4.13 \pm  1.71$ & $ 1.49 \pm  0.97$ & \dots & \dots \\
OGLE-mNOVA-27 & $-0.33 \pm  1.44$ & $ 3.21 \pm  2.41$ & \dots & \dots \\
OGLE-mNOVA-28 & $-1.22 \pm  2.05$ & $ 0.05 \pm  1.94$ & \dots & \dots \\
OGLE-mNOVA-29 & $-1.85 \pm  2.75$ & $-0.54 \pm  2.33$ & \dots & \dots \\
\enddata
\tablecomments{$^{(a)}$The \textit{Gaia} DR3 \citep{gaia2016,gaia2023} astrometric solution is unreliable because the Renormalized Unit Weight Error is larger than 1.4 \citep{lindegren2021}.}
\end{deluxetable*}

\onecolumngrid

\restartappendixnumbering

\section{\lowercase{e}ROSITA Upper Limits}
\label{app:erosita}

The first catalog of X-ray sources detected by the eROSITA telescope array \citep{predehl2021} on board the Spectrum-Roentgen-Gamma (SRG; \citealt{sunyaev2021}) satellite was recently published by \citet{merloni2024}. The number of exposures, and so the depth of the survey, depends on sky position. Thanks to the SRG scanning law, objects located near the ecliptic poles are observed most frequently. Coincidentally, the LMC is located near the south ecliptic pole.

We cross-matched our list with the eROSITA catalog \citep{merloni2024}, but none of the objects from Table~\ref{tab:list} were detected by eROSITA. The catalog of \citet{merloni2024} contains X-ray sources located in the western Galactic hemisphere detected during the first six months of eROSITA operations (from 2019 December 12 to 2020 June 11, that is, $2,458,829\! <\! \mathrm{JD}\! <\! 2,459,011$). The overlap with OGLE observations is relatively small because the OGLE-IV operations were suspended on 2020 March 18 ($\mathrm{JD}=2,458,927$) due to the COVID-19 pandemic. Nonetheless, we found that four objects (OGLE-mNOVA-06, OGLE-mNOVA-10, OGLE-mNOVA-19, and OGLE-mNOVA-24) were in outburst during the first eROSITA All-Sky Survey. All four objects were relatively faint and reached a peak $I \sim 19.55\!-\!19.98$ during that time. Assuming that the X-ray flux scales proportionally with peak $I$-band flux, and taking OGLE-mNOVA-11 as a benchmark, the maximal expected X-ray flux is $\sim (6\!-\!8)\times 10^{-14}$ erg\,cm$^{-2}$\,s$^{-1}$ in the 0.3--10\,keV range. The expected flux in the eROSITA soft and medium energy bands (0.2--2.3\,keV) is about 6\% larger at $(6\!-\!9)\times 10^{-14}$ erg\,cm$^{-2}$\,s$^{-1}$. However, each object must have been scanned by eROSITA multiple times, and the time-averaged flux is likely to be smaller.

We queried the eROSITA upper limits service \citep{tubin_arenas_2024} to get upper flux limits in the 0.2--2.3 keV range for all four objects discussed above. These limits are calculated at a 99.87\% ($3\sigma$) one-sided confidence level, assuming a power-law spectrum with a photon index $\Gamma=2.0$ and a column density $N_{\rm H}=3 \times 10^{20}$\,cm$^{-2}$. We obtained limits of $(5.1, 1.9, 2.0, 3.8)\times 10^{-14}$ erg\,cm$^{-2}$\,s$^{-1}$ at the positions of OGLE-mNOVA-(06,10,19,24), respectively. We followed a procedure outlined in \citet{tubin_arenas_2024} to estimate upper limits for the best-fitting absorbed blackbody model presented in Appendix~\ref{app:swift}. Using the appropriate eROSITA calibration files and \textsc{XSPEC} version~12.14.0 \citep{arnaud1996}, we estimated the energy-to-count conversion factor of $1.037\times 10^{12}$\,cm$^2$\,erg$^{-1}$, which is only 3.4\% smaller than that calculated using the fiducial model in \citet{tubin_arenas_2024}. Therefore, the upper limits for our best-fitting model are only 3.4\% larger than default ones.
In the 0.2--5 keV range, the corresponding limits are $(7.5, 3.7, 2.5, 5.4)\times 10^{-14}$ erg\,cm$^{-2}$\,s$^{-1}$. We thus conclude that the expected time-averaged flux is close to or below the current eROSITA limits. It is therefore possible that the selected stars may be detected in future eROSITA data releases once more data are processed or may be detected in data from individual scans.

\bibliography{pap}{}
\bibliographystyle{aasjournal}

\end{document}